\documentstyle[12pt]{article}
\def\a{\alpha}\def\b{\beta}\def\g{\gamma}\def\d{\delta}
\def\E{\varepsilon}
\def\l{\lambda}\def\s{\sigma}
\def\om{\omega}\def\G{\Gamma}
\newcommand{\p}[1]{(\ref{#1})}

\begin{document}
\renewcommand{\thefootnote}{\fnsymbol{footnote}}
\thispagestyle{empty}
\def\theequation{\thesection.\arabic{equation}}

\begin{flushright}
{\bf ITP-UWr-921-98 \\
TUW-98-23 \\
hep-th/9811022} \\
 October 27, 1998 \\
misprints corrected January 1999. 
\end{flushright}

\vskip2cm

\begin{center}
{\Large \bf Tensorial Central Charges
and New Superparticle Models with Fundamental Spinor Coordinates}

\bigskip

\bigskip

\bigskip

{\large \bf Igor Bandos$^{1)}$\footnote{ Lise Meitner Fellow.
On leave of absence from  Institute for Theoretical Physics,
NSC Kharkov Institute of Physics and Technology,
310108 Kharkov, Ukraine}
 and Jerzy Lukierski}$^{2)}$\footnote{Supported in part by {\bf KBN} grant
{\bf 2P03B13012}}
 \\
$^{1)}$ {\it Institute for Theoretical Physics\\
Technical University of Vienna\\
Wiedner Haupstrasse 8-10\\
A-1040 Wien, Austria}\\
{\bf e-mail: bandos@tph32.tuwien.ac.at}
 \\
$^{2)}${\it Institute for Theoretical Physics
\\
University of Wroclaw\\
50-204 Wroclaw, Poland}
\\
{\bf e-mail: lukier@proton.ift.uni.wroc.pl}
\date{}
\end{center}

\vskip2cm


\begin{abstract}

We consider firstly  simple $D=4$ superalgebra with six real tensorial central charges
$Z_{\mu\nu}$, and discuss its possible realizations in massive and
massless cases. Massless case is dynamically realized by generalized
Ferber-Shirafuji (FS) model with fundamental
bosonic spinor coordinates.
The Lorentz invariance is not broken due to the realization of central charges
generators in terms of bosonic spinors.
The model contains four fermionic
coordinates and possesses three $\kappa$-symmetries thus providing
the BPS configuration preserving $3/4$ of the target space supersymmetries.
We show that the
physical degrees of freedom ($8$ real bosonic and $1$ real
Grassmann variable) of our
model can be described by $OSp(8|1)$ supertwistor.
 The relation with recent superparticle model by Rudychev and Sezgin is pointed out.
 Finally
we propose  a higher dimensional generalization of our model
with one real fundamental bosonic spinor.
$D=10$  model describes massless superparticle with
composite tensorial central charges and in $D=11$ we obtain 0-superbrane
model with  nonvanishing mass which is generated dynamically.

\end{abstract}

\setcounter{page}0
\renewcommand{\thefootnote}{\arabic{footnote}}
\setcounter{footnote}{0}
\newpage

\section{Introduction}
\setcounter{equation}{0}

Recently it became clear that in supersymmetric theories besides
scalar central charges,
 which are present in conventional D=4 Poincar\'{e}
supersymmetry scheme of Haag, Lopusza\'{n}ski and Sohnius \cite{HLS}
one should consider also nonscalar generalized central charges: - tensorial
\cite{HP}--\cite{Hewson}  or even
spinor ones \cite{AF,S}.
Such generalized central extension of  standard
$N=1$ $D$--dimensional supersymmetry algebra
$\{ Q_{\alpha}, Q_{\beta } \} = i \Gamma^m _{\alpha \beta} P_m $
can be written in the form
\begin{equation}\label{QQZ}
\{ Q_{\alpha}, Q_{\beta} \} = Z_{\alpha\beta} ,
\end{equation}
where $Z_{\alpha\beta}$ is the most general
 symmetric matrix
 of Abelian generalized central charges.

The tensorial central charges $Z^{m_1\ldots m_p}$
 appear through decomposition of
 the symmetric matrix $Z_{\alpha\beta}$ on the  basis defined
 by the products $\Gamma^{(p)} \equiv \Gamma^{m_1 \ldots m_p} =
 \Gamma^{[m_1}\ldots\Gamma^{m_p]}$ of $D$-dimensional $\Gamma^m$ matrices.
\begin{equation}\label{Zdec}
Z_{\alpha\beta} =
\left( \Gamma_{m} 
C\right)_{(\alpha\beta )} P^{m} +
\sum\limits_{symmetric} \left( 
\Gamma_{m_{1} \ldots m_{p}} C \right)_{( \alpha\beta )} Z^{m_{1} \ldots
m_{p}}\, , \end{equation}
It should be noted that, as only symmetric matrices are
really involved on the right hand side of \p{Zdec}, for particular dimensions
and signatures one can also consider the superalgebras \p{Zdec} without the momenta
$P_{m}$ described by the term linear in $\Gamma_m$.

Main aim of this paper is to discuss the appearance of the tensorial central charges in the case of 'physical'
$D=4$ ($D=1+3$) supersymmetry.
If, for simplicity, we consider $N=1$
supersymmetry, one can generalize the standard $D=4$ superalgebra as
follows
\begin{equation}\label{QQZ4}
\{ Q_A, Q_B \} = Z_{AB} , \qquad
\{ \bar{Q}_{\dot{A}}, \bar{Q}_{\dot{B}} \} = \bar{Z}_{\dot{A}\dot{B}} , \qquad
\end{equation}
$$
\{ Q_A, \bar{Q}_{\dot{B}} \} = P_{A\dot{B}} , \qquad
$$
where $ (Q_A)^* = \bar{Q}_{\dot{A}}$,
$(P_{A\dot{B}})^* = P_{B\dot{A}}$,
$(Z_{{A}{B}})^* =
\bar{Z}_{\dot{A}\dot{B}}$ and six real
commuting central charges $Z_{\mu\nu}= - Z_{\nu\mu}$ are related to the
symmetric complex
spin-tensor $Z_{AB}$ by
\footnote{For two-component $D=4$ Weyl spinor formalism see e.g. \cite{OM75}.
We have \\ $(\s_{mn})_A^{~B}= {1\over 2i} \big(
(\s_{\mu} )_{A\dot{B}}
\tilde{\s}_{\nu}^{\dot{B}B}-
(\s_{\nu} )_{A\dot{B}}
\tilde{\s}_{\mu}^{\dot{B}B}\big)=
 - {i \over 2} \E_{\mu\nu\rho\l}(\s^{\rho\l})_A^{~B} =
[(\tilde{\s}_{\mu\nu})^{\dot{B}}_{~\dot{A}}]^* $. }
\begin{equation}\label{ZZZ}
Z_{\mu\nu} = {i \over 2}\left( \bar{Z}_{\dot{A}\dot{B}}
\tilde{\s}_{\mu\nu}^{\dot{A}\dot{B}}
- Z_{AB} \s_{\mu\nu}^{AB}\right).
\end{equation}
Thus the spin-tensors $Z_{AB}$ and $Z_{\dot{A}\dot{B}}$
$$
Z_{AB} ={i \over 4} Z_{\mu\nu} \s^{\mu\nu}_{AB}, \qquad
\bar{Z}_{\dot{A}\dot{B}} =
 - {i \over 4 }{Z}_{\mu\nu} \tilde{\s}^{\mu\nu}_{\dot{A}\dot{B}}
$$
represent the self-dual and anti-self-dual parts of the central charge
matrices.

The tensorial central charges \p{ZZZ} commute with fourmomenta
$P_{\mu} = {1 \over 2} \tilde{\s}_{\mu}^{\dot{B}A} P_{A\dot{B}}$ and
transform as a tensor under the Lorentz group with generators
$M_{\mu\nu}=-M_{\nu\mu}$  ($\eta_{\mu\nu} = diag (1, -1, -1, -1)$)
\begin{equation}\label{MZZ}
[ M_{\mu\nu}, Z_{\rho\l}]  = -i (\eta_{\mu\rho} Z_{\nu\l}
- \eta_{\nu\rho} Z_{\mu\l} -
\eta_{\mu\l} Z_{\nu\rho} + \eta_{\nu\l} Z_{\mu\rho} ).
\end{equation}

In this paper we shall consider
two aspects of the appearance of tensorial central charges in
$D=4$ superalgebra:

{\bf i.}
 Their impact on the superalgebraic representations -- in Section {\bf 2}. We consider separately massive
 ($P_\mu P^\mu > 0$) and massless ($P_\mu P^\mu=0$) cases. We recall that
 in the massless case half of the fermionic degrees of freedom can be
 eliminated
what leads to the shortening of the massless supermultiplets and, hence,
 only $N=1/2$ supersymmetries acts nontrivially.
It appears that the presence of particular form of the
central charge
\begin{equation}\label{ZKP}
Z_{\mu\nu} = K_{\mu} P_{\nu} - K_{\nu} P_{\mu}
\end{equation}
provides the additional shortening of the
massless supermultiplet  (with $N=1/4$ supersymmetry realized nontrivially).

{\bf ii.}  Their dynamical consequences -- in Section {\bf 3}.
The formula \p{ZKP} for tensorial central charge can be derived
from the  generalized Ferber--Shirafuji (FS) model
\cite{F78,S83} with fundamental spinor coordinates
$\l_\a = (\l_A, \l_{\dot{A}})$ and additional central charge coordinates
$z^{AB}=z^{BA} = [z^{\dot{A}\dot{B}}]^*$.
In such a model three Grassmann degrees of freedom
 out of four are pure gauge and can be gauged away by $\kappa$--transformations
 \cite{AL}.
In the language of brane physics
\cite{PKT,Bars,Hewson,SorT} such model
corresponds to BPS configuration preserving
$3/4$ of the target space supersymmetry.
Such configurations were not known before.

The model can be reformulated in terms of two Weyl spinors
$\l_A, \mu_A$ and one real Grassmann variable $\zeta$
expressed by the
generalization
of supersymmetric  Penrose--Ferber
relations \cite{F78,S83,BBCL} between supertwistor and superspace coordinates.
  Such reformulation is described by $OSp(8|1)$ invariant free supertwistor
  model with the action
\begin{equation}
 \label{act0}
S = - {1 \over 2} \int d \tau {Y}_{{\cal A}} G^{{\cal A} {\cal B}}
\dot{Y}_{{\cal B}}
\end{equation}
where ${Y}_{{\cal A}} = (y_1, \ldots , y_8; \zeta)
\equiv
(\l_\a , \mu^\a, \zeta )$ is the real $SO(8|1)$ supertwistor
  (see e.g. \cite{Lstw}) and
\begin{equation}
 \label{OSpm}
  G^{{\cal A}{\cal B}} = \left( \matrix{
   \om^{(8)} & 0 \cr
   0 & 2i \cr} \right) =
   \left(
  \matrix{
  \left(\underline{\matrix{
   0_2 & I_2 & 0_2 & 0_2  \cr
  -I_2 & 0_2 & 0_2 & 0_2   \cr
   0_2 & 0_2 & 0_2 & I_2   \cr
  0_2 & 0_2 & {-I}_2 & 0_2 \cr  }}\right)
    & |& 0   \cr
   0 & |& 2i \cr}           \right)
\end{equation}
 is the $OSp(8|1)$ supersymplectic structure with bosonic $Sp(8)$ symplectic
 metric \\ $\om^{(8)}=-(\om^{(8)})^T$.
It should be mentioned therefore that due to the presence of
 tensorial central charges the standard $SU(2,2|1)$ supertwistor description
 \cite{F78,S83,BBCL,stv,SorG,Town91}
 of the Brink--Schwarz (BS) massless superparticle \cite{BS}
 with one complex Grassmann coordinate
 is replaced by a model with $OSp(8|1)$ invariance and one real Grassmann degree of
 freedom.

It should be stressed that by the use of spinor coordinates in the presence of tensorial central charges

\begin{itemize}
\item
we do not increase the initial number of spinor degrees of freedom
(four complex or eight real components)
in comparison with the model without
tensorial central charges;
\item we keep the manifest Lorentz invariance despite the presence of
tensorial central charges.
\end{itemize}

In fact, when we use our formulae (see Section {\bf 3})

 \begin{equation}\label{CPrep}
P_{A\dot{B}} = \lambda_{A}\bar{\lambda}_{\dot{B}},
\qquad
Z_{AB} = \lambda_{A} \lambda_{B},
\qquad
\bar{Z}_{\dot{A}\dot{B}} = \bar{\lambda}_{\dot{A}}
\bar{\lambda}_{\dot{B}}
\end{equation}
we find
 that, in comparison with standard FS model
 ($P_{A\dot{B}} = \l_A\bar{\l}_{\dot{B}}$, 
$Z_{AB}=\bar{Z}_{\dot{A}\dot{B}}=0$),
 only the phase of
 spinor $\l_A$ becomes an additional physical bosonic degree of freedom.

We also show that our model can be related with generalized superparticle model of Rudychev
and Sezgin \cite{RS}.
In Section {\bf 4} we will describe the Rudychev--Sezgin model
for $D=1+3$ and
find the general solution of the BPS constraint \cite{RS} in
terms of two bosonic spinors
$(\l_A, \mu_A)$. It appears that by putting $u_A=0$ and fixing
one normalization factor we arrive at our model.

In Section {\bf 5}
we propose a  generalization of our model for $D>4$
with one real fundamental spinor.
In $D=10$ the model describes a massless superparticle with composite
tensorial central charges. In $D=11$ we get the 0-superbrane model with mass
 generated dynamically in a way analogous to the brane tension generation
\cite{generation}.

In Section {\bf 6}
we present final  remarks.

\bigskip

\section{On representations  of $N=1$, $D=4$ superalgebra
with tensorial central charge}
\setcounter{equation}{0}

In order to describe the supersymmetry multiplets for the algebra
\p{QQZ4} we shall consider supercharges
$Q_A$, $\bar{Q}_{\dot{A}}$ in a particular Lorentz frame.
We shall consider separately the massive $P_\mu P^\mu = M^2 > 0$
 and massless $P_\mu P^\mu  = 0$ cases.

\subsection*{ $ M^2 > 0$:}

We choose the rest frame for the fourmomentum, i.e.
$P_\mu =(M,0,0,0)$.
In such a way we obtain the algebra \p{QQZ4} in the following
$U(2)$ invariant form
 \begin{equation}\label{QQPZ1}
\{ Q_A, {Q}^{\dagger}_{\dot{B}} \} = M \d_{A\dot{B}} , \qquad
\end{equation}
Further we use the $U(2)$ 
transformations (space rotations $SO(3)=SU(2)$ plus internal $U(1)$)
to transform the central charge matrices to the form ($U \in U(2)$; see
\cite{Zumino})
 \begin{equation}\label{Zdiag}
Z= UZ^{(0)}U^+ , \qquad Z^{(0)} = \left( \matrix{ a & 0 \cr
                                                   0 & b \cr}\right)
= x I + y \s_3
\end{equation}
where $a$ and $b$ are real and positive, $x={1 \over 2} (a+b)$,
$y={1 \over 2} (a-b)$.

Introducing the fourdimensional Majorana spinor
 \begin{equation}\label{QMj}
Q_\a = \left( \matrix{ Q_A \cr
                        {Q}^{\dagger}_{\dot{A}}    \cr}\right)
\end{equation}
one obtains for the matrix of commutator of supercharges
 \begin{equation}\label{SQQ}
S_{\a\b} = \{ Q_\a , Q_\b \}
= \left( \matrix{ xI_2+y\s_3 & M~I_2 \cr
                    M~I_2 & xI_2 - y \s_3 \cr}\right)
\end{equation}
with
 \begin{equation}\label{detS}
det~S~ =  (M^2 - a^2)~(M^2 - b^2), \qquad
\end{equation}
The matrix $S$ can be diagonalized and
the number of
supersymmetries acting nontrivially
corresponds to the number of nonvanishing diagonal elements of the
matrix $S$ (compare e.g. with \cite{FSZ}).
Thus we should  consider the following three cases:
\begin{enumerate}
\item If $M \not= \pm a$ and $M \not= \pm b$ there are four nontrivial
supersymmetries. The diagonal form of the matrix \p{SQQ} is
$diag (a+M, a-M, b+M, b-M)$.
\item If $M= \pm a$ or $M=\pm b$, but $a \not= b$, then one of the
eigenvalues is equal to zero.
Denoting the corresponding charge by $Q_{null}$ one gets
$\{ Q_{null}, Q^+_{null} \} = 0$ and,
if $Q_{null}^+ | 0 > = 0$ and $ Q_{null} | 0 > = | 1 >$, one obtains that
$|<1|1>|^2 = 0$. Assuming that the representation space of our superalgebra
is span by
the positive norm states, we should discard $Q_{null}$ as generating trivial representations.
\item If $a=b$ and $m=\pm a$ only two supercharges generate nontrivial
representations,
i.e. we obtain only $N={1 \over 2}$ $D=4$ supersymmetries.
\end{enumerate}

\subsection*{$M=0$: }

In such a case, using the light--cone frame
$P_m =(p,0,0, p)$ ($p_0 =|${\bf{p}}$|=p$ one gets)
 \begin{equation}\label{QQP0}
\{ Q_A, {Q}^\dagger_{\dot{B}} \} = \d_{A1} \d_{\dot{B}1} 2p ~, \qquad
\end{equation}
In our framework the supercharges $Q_2$ generate trivial representation space.
If we assume that does exist a nontrivial Clifford vacuum
($ Q_2^\dagger |0>_2 = 0,  ~~{}_2<0|0>_2 =1,$)
then we obtain
from
${}_2<0|Q^\dagger_2| 0>_2 =
\bar{Z}_{22}{~~}_2<0| 0>_2 = 0 $ that $\bar{Z}_{22}=Z_{22}=0$.

We shall  assume further that
 \begin{equation}\label{ZZ=0}
 Z_{AB} Z^{AB} = \bar{Z}_{\dot{A}\dot{B}}
\bar{Z}^{\dot{A}\dot{B}} = 0,
\end{equation}
what implies $Z_{12}=0$. Thus we arrive at the algebra with
all nontrivial relations being collected in $Q_1, {Q}^\dagger_1$ sector
 \begin{equation}\label{Q1Q1}
\{ Q_1, {Q}^\dagger_1 \} = 2p , \qquad
Q_1^2 = Z_{11}, \qquad {Q}_1^{\dagger 2} = \bar{Z}_{\dot{1}\dot{1}} .\qquad
\end{equation}
Because the relations \p{Q1Q1} are invariant under
the phase transformations $Q_1 \rightarrow e^{i\a}Q_1$,
$Z_{11} \rightarrow e^{2i\a} Z_{11}$, one can
fix the central charge $Z_{11}$ to be real
$Z_{11} = \bar{Z}_{\dot{1}\dot{1}}= r$.
Introducing
 \begin{equation}\label{Rpm}
R_+={1 \over 2} (Q_1 + {Q}^+_1 ), \qquad
R_-={1 \over 2} (Q_1 - {Q}^+_1 ), \qquad
\end{equation}
one gets
 \begin{equation}\label{RR}
\{ R_+, R_+ \} = M_+ = (p+r), \qquad
\{ R_-, R_- \} = M_- = (p-r), \qquad
\end{equation}
$$
\{ R_+, R_- \} = 0.
$$

We can distinguish the following two cases:
\begin{enumerate}
\item $r \not= p$. In such a case we have two nontrivial supersymmetries
(as in the case $r=0$).
\item
$r = \pm p $. In such a case {\sl remains only one nontrivial supersymmetry}.
Such a case can be described in covariant way by the spinor ansatz
\p{CPrep} \footnote{Gauge fixing  corresponding to \p{Q1Q1} is given by
$$\l_A = \sqrt{2p} \left( \matrix{1 \cr 0\cr }\right)$$}. It can be shown that the corresponding
tensorial central charge can be expressed by Eq. \p{ZKP} where
\begin{equation}\label{KKs}
K_m = {1 \over 2} \tilde{\s}_m^{\dot{B}A}
K_{A\dot{B}}, \qquad
K_{A\dot{B}} = 2 (\l_A \hat{\bar{\mu}}_{\dot{B}} + \hat{\mu}_A \l_{\dot{B}} )
\end{equation}
and $\l^A \hat{\mu}_A = 1$. In our special coordinate frame we should
choose $K_m = (0,1,0,0)$.
\end{enumerate}


We see therefore that the presence of tensorial central charges can reduce the
number of nontrivial supersymmetries to one.
In the dynamical model this should be realized by the presence of
additional third $\kappa$--symmetry.
In the next section we consider the relations \p{CPrep}, and, thus,
\p{ZKP} and \p{KKs},
built in as the dynamical constraints.

\section{Generalization of Ferber--Shirafuji superparticle model:
spinor fundamental variables and central charges}
\setcounter{equation}{0}

 We generalize the model presented in \cite{S83} as follows
\begin{equation}\label{action}
S = \int d \tau
\left( \lambda_{A}\bar{\lambda}_{\dot{B}} \Pi_\tau^{A\dot{B}} +
\lambda_{A}\lambda_{B}  \Pi_\tau^{AB}
+ \bar{\lambda}_{\dot{A}} \bar{\lambda}_{\dot{B}}
\, \Pi_\tau^{\dot{A}\dot{B}} \right)\, ,
\end{equation}
where
\begin{equation}\label{vielbeine}
\begin{array}{l}
\displaystyle
\Pi^{A\dot{B}} \equiv  d\tau \Pi_\tau^{A\dot{B}}
= d{X}^{A\dot{B}}
+ i \left( d\Theta^{A} \bar{\Theta}^{\dot{B}}
- \Theta^{A} d\bar{\Theta}^{\dot{B}}\right) \, ,
\\ \nonumber
\displaystyle
\Pi^{AB} \equiv d\tau \Pi_\tau^{AB} =
d{z}^{AB} - ~i~ \Theta^{(A}~d{\Theta}^{B)}\, ,
\\  
\displaystyle
\bar{\Pi}^{\dot{A}\dot{B}}
\equiv d\tau \bar{\Pi}_\tau^{\dot{A}\dot{B}}
= d{\bar{z}}^{\dot{A}\dot{B}}
- ~i ~\bar{\Theta}^{(\dot{A}} ~d{\bar{\Theta}}\,{}^{\dot{B})}\, ,
\end{array}
\end{equation}
with 
$$
d\Theta^{(A} \Theta^{B)}={1 \over 2} (d\Theta^{A} \Theta^{B} + 
d\Theta^{B} \Theta^{A}) 
$$
are the supercovariant one--forms in $D=4$, $N=1$ generalized flat superspace
\begin{equation}\label{superspace}
M^{(4+6|4)} = \{ Y^M \} \equiv \{ (x^{A\dot{A}}, z^{AB},
\bar{z}^{\dot{A}\dot{B}}; \Theta^{A}, \bar{\Theta}^{\dot{A}})\},
\end{equation}
{\sl with tensorial central charge coordinates}
 $z^{mn} = (z^{AB}, \bar{z}^{\dot{A}\dot{B}} )$ (see \p{ZZZ}).
The complete configuration space of the model \p{action}  contains
 additionally the complex-conjugate pair $(\l_A, \bar{\l}_{\dot{A}})$
 of Weyl spinors                      
\begin{equation}\label{confsuperspace}
{\cal M}^{(4+6+4|4)} = \{ q^{{\cal M}} \} \equiv
\{  (Y^M; \l^A, \bar{\l}^{\dot{A}}) \}
=\{ (x^{A\dot{A}}, z^{AB},
\bar{z}^{\dot{A}\dot{B}};  \l^A, \bar{\l}^{\dot{A}};
\Theta^{A}, \bar{\Theta}^{\dot{A}})\},
\end{equation}

Calculating the canonical momenta
\begin{equation}\label{momenta}
{\cal P}_{{\cal M}} = { \partial L \over \partial \dot{q}^{{\cal M}} }
= (P_{A\dot{A}}, Z_{AB},
\bar{Z}_{\dot{A}\dot{B}};  P^A, \bar{P}^{\dot{A}};
\pi^{A}, \bar{\pi}^{\dot{A}}) ,
\end{equation}
we obtain the following set of the
primary constraints

\begin{equation}\label{Phi1}
\Phi_{A\dot{B}} \equiv
P_{A\dot{B}} - \lambda_{A} \bar{\lambda}_{\dot{B}} = 0 ,
\end{equation}
\begin{equation}\label{Phi2}
\Phi_{A{B}} \equiv
Z_{A{B}} - \lambda_{A} {\lambda}_{{B}} = 0 ,
\end{equation}
\begin{equation}\label{Phi3}
\Phi_{\dot{A}\dot{B}} \equiv
\bar{Z}_{\dot{A}\dot{B}} - \bar{\lambda}_{\dot{A}} \bar{\lambda}_{\dot{B}} = 0 ,
\end{equation}
\begin{equation}\label{PA=0}
P_{A}=0, \qquad
\bar{P}_{\dot{A}}  = 0 ,
\end{equation}
                  
\begin{equation}\label{DA=0}
D_{A}\equiv - \pi_A + i P_{A\dot{B}} \bar{\Theta}^{\dot{B}} + i
Z_{AB} \Theta^B=0, \qquad
\end{equation}
\begin{equation}\label{bDA=0}
\bar{D}_{\dot{A}}\equiv  \bar{\pi} _{\dot{A}} - i \Theta^B  P_{B\dot{A}}
- i \bar{Z}_{\dot{A}\dot{B}} \bar{\Theta}^{\dot{B}} = 0.
\end{equation}

Because the action \p{action} is invariant under the world line
reparametrization, the canonical Hamiltonian vanishes
 \begin{equation}\label{H=0}
H \equiv \dot{q}^{{\cal M}}{{\cal P}}_{{\cal M}} - L (q^{{\cal M}} ,
\dot{q}^{{\cal M}}) = 0
\end{equation}
It can be deduced that  the set \p{Phi1}-\p{bDA=0} of $14$
bosonic and $4$ fermionic constraints contains
$6$ bosonic and $3$ fermionic first class constraints
\begin{equation}\label{B1}
 B_1 =
 {\l}^{A} {\bar{\l}}^{\dot{B}}
 P_{A\dot{B}}   = 0,
\end{equation}
\begin{equation}\label{B2}
 B_2 =
 {\l}^{A} \hat{\bar{\mu}}^{\dot{B}}
 P_{A\dot{B}}  - \l^A \hat{\mu}^B Z_{AB} = 0,
\end{equation}
\begin{equation}\label{B3}
 B_3 \equiv (B_2)^*=
 \hat{\mu}^{A} {\bar{\l}}^{\dot{B}}
 P_{A\dot{B}}  - \bar{\l}^{\dot{A}} \hat{\bar{\mu}}^{\dot{B}}
 \bar{Z}_{\dot{A}\dot{B}} = 0,
\end{equation}
\begin{equation}\label{B4}
 B_4 =
2 \hat{\mu}^{A} \hat{\bar{\mu}}^{\dot{B}}
 P_{A\dot{B}}  - \hat{\mu}^A \hat{\mu}^B Z_{AB}
 - \hat{\bar{\mu}}^{\dot{A}} \hat{\bar{\mu}}^{\dot{B}} \bar{Z}_{\dot{A}\dot{B}}
  = 0,
\end{equation}
\begin{equation}\label{B5}
 B_5 =
 {\l}^{A} {\bar{\l}}^{{B}}
 Z_{AB} = 0,
\end{equation}
\begin{equation}\label{B6}
 B_6 \equiv (B_5)^*=
 \bar{\l}^{\dot{A}} {\bar{\l}}^{\dot{B}}
 \bar{Z}_{\dot{A}\dot{B}} = 0,
\end{equation}
\begin{equation}\label{F1}
 F_1 =
 {\l}^{A}
 D_{A} = 0,
\end{equation}
\begin{equation}\label{F2}
 F_2 \equiv (F_1)^*=
 \bar{\l}^{\dot{A}} {\bar{D}}_{\dot{A}}= 0,
\end{equation}
\begin{equation}\label{F3}
 F_3 =
 \hat{\mu}^{A}
 D_{A} +
 \hat{\bar{\mu}}^{\dot{A}} {\bar{D}}_{\dot{A}}= 0,
\end{equation}
where we assume that $\l^A \mu_A \not= 0$ and
\begin{equation}\label{hatmu}
 \hat{\mu}^{A}= {\mu^A \over \l^B \mu_B} , \qquad
 \hat{\bar{\mu}}^{\dot{A}} =
{ \bar{\mu}^{\dot{A}} \over \l^{\dot{B}}
{\mu}_{\dot{B}}},
\end{equation}
i.e. $\l^A \hat{\mu}_A = \bar{\l}^{\dot{A}} \hat{\bar{\mu}} =1$.
One can show
\footnote{
We recall  \cite{Dirac} that the first
class constraints are defined as
those whose
Poisson brackets with all constraints weakly vanish.
Then one can show \cite{Dirac}  that the first class constraints  form
the closed
algebra. }
that our
first class constraints \p{B1} - \p{F3}
can be chosen  for any particular form of the
second spinor $\mu^A$ as a function of canonical
variables $(q^{\cal M}, {\cal P}_{{\cal M}})$.
Further we shall propose and motivate the choice
for $\mu^A, ~\bar{\mu}^{\dot{A}}$.

The remaining
$8$ bosonic and $1$ fermionic constraints are the second class ones.
They are
\begin{equation}\label{SB12}
 {\l}^{A} \hat{\bar{\mu}}^{\dot{B}}
 P_{A\dot{B}}  + \l^A \hat{\mu}^B Z_{AB} = 0,  \qquad
 \hat{\mu}^{A} {\bar{\l}}^{\dot{B}}
 P_{A\dot{B}}  + \bar{\l}^{\dot{A}} \hat{\bar{\mu}}^{\dot{B}}
 \bar{Z}_{\dot{A}\dot{B}} = 0,
\end{equation}
\begin{equation}\label{SB34}
 \hat{\mu}^A \hat{\mu}^B Z_{AB} - 1 = 0, \qquad
\hat{\bar{\mu}}^{\dot{A}} \hat{\bar{\mu}}^{\dot{B}}
\bar{Z}_{\dot{A}\dot{B}} -1
  = 0,
\end{equation}
\begin{equation}\label{SB56}
 P_{A} =0, \qquad
 \bar{P}_{\dot{A}} = 0,
\end{equation}

\begin{equation}\label{SF1}
 S_F \equiv \hat{\mu}^{A}
 D_{A} -
 \hat{\bar{\mu}}^{\dot{A}} {\bar{D}}_{\dot{A}}= 0,
\end{equation}
 We see that the number $\#$ of on-shell phase space degrees of freedom
 in our model is
\begin{equation}
 \label{number}
\# = (28_B +8_F) - 2 \times (6_B + 3_F) - (8_B + 1_F) = 8_B + 1_F
\end{equation}
in distinction with the standard massless superparticle model of
Brink--Schwarz
\cite{BS} or Ferber-Shirafuji \cite{F78,S83} containing $6_B + 2_F$
physical degrees of freedom.

In order to explain the difference
in the number of fermionic constraints, let us write down the matrices of
Poisson brackets
for the fermionic constraints \p{DA=0}, \p{bDA=0}.
 In our case it has the form
\begin{equation}
 \label{CDD}
   C_{\a\b} = \left( \matrix{
    \{ D_{A}, D_{B} \}_P  &  \{ D_{A}, \bar{D}_{\dot{B}} \}_P   \cr
     \{ \bar{D}_{\dot{A}}, D_{{B}} \}_P &
 \{ \bar{D}_{\dot{A}}, \bar{D}_{\dot{B}} \}_P \cr }\right)
  = 2i  \left( \matrix{
    \l_{A}\l_{B} &  \l_{A} \bar{\l}_{\dot{B}}   \cr
     \bar{\l}_{\dot{A}} \l_{{B}}  &
 \bar{\l}_{\dot{A}} \bar{\l}_{\dot{B}} \cr }\right)
\end{equation}
while for the standard FS model \cite{F78,S83}
we obtain
\begin{equation}
 \label{FSCDD}
   C^{FS}_{\a\b}
  =  2i \left( \matrix{
            0 &  \l_{A} \bar{\l}_{\dot{B}}   \cr
     \bar{\l}_{\dot{A}} \l_{{B}}  & 0
 \cr }\right)
\end{equation}
Now it is evident that in our case the rank of the matrix $C$ is one,
while for FS model it is equal to two
$$
rank (C) = 1, \qquad
  rank( C^{FS}) =2.
$$
Consequently,  in our model  there are three fermionic first class constraints
generating three $\kappa$--symmetries, one more than in the FS model.

\bigskip

In order to clarify the meaning of the superparticle model \p{action}
and present an explicit representation for its physical degrees of freedom,
we shall demonstrate that it admits the supertwistor representation
in terms of independent bosonic spinor $\l^A$, bosonic spinor $\mu^A$
being composed of $\l^A$ and superspace variables
\begin{equation}
 \label{muXZ2}
 \mu^{A} =
 \left(  x^{A\dot{B}}
 + i \Theta^{A} \bar{\Theta}^{\dot{B}}
 \right)
  \bar{\lambda}{}_{{\dot{B}}}
+ 2 z^{AB} \lambda_{B}
+ i \Theta_{A} (\Theta^B
  \lambda_B ), 
\end{equation}
\begin{equation}
 \label{bmuXZ2}
 \bar{\mu}^{\dot{A}} =
 \left(  x^{B\dot{A}}
 - i \Theta^{B} \bar{\Theta}^{\dot{A}}
 \right) 
  \lambda_{B}
+ 2\bar{z}^{\dot{A}\dot{B}} \bar{\lambda}_{\dot{B}}
- i \bar{\Theta}^{\dot{A}}
\bar{\Theta}^{\dot{B}}
  \bar{\lambda}_{\dot{B}}
\end{equation}
 and one real  fermionic composite Grassmann variable $\zeta$
\begin{equation}
 \label{zetaTh2}
\zeta = \Theta^A
{\lambda}_{A}
+
\bar{\Theta}^{\dot{A}}
  \bar{\lambda}_{\dot{A}}
\end{equation}
 Eqs. \p{muXZ2} -\p{zetaTh2} describe  $OSp(8|1)$--supersymmetric
 generalization
 of the Penrose correspondence which is alternative
 to the previously known $SU(2,2|1)$ correspondence, firstly proposed by Ferber \cite{F78}.
 Performing integration by parts and neglecting boundary terms we
can express  our action \p{action}  in terms of
$OSp(8|1)$ supertwistor variables as follows:
\begin{equation}\label{actiontw}
 S = - \int  \left(
 {\mu}{}^{A}
d\lambda_{A}
  + \bar{\mu}{}^{\dot{A}} \,
 d{\bar{\lambda}}_{\dot{A}}
  + i d{\zeta} ~{\zeta} \right) \, .
\end{equation}
Eq. \p{actiontw} presents  the free $OSp(8|1)$  supertwistor
action.
 It can be rewritten in the form \p{act0}
with real coordinates
$Y^{A} = ( \mu^\a , \l^\a, \zeta)$ where
real Majorana spinors $\mu^\a, \l^\a$ are obtained from the
Weyl spinors
$(\mu^A, \bar{\mu}^{\dot{A}})$, $(\l^A, \bar{\l}^{\dot{A}})$
by a linear transformation changing
for the $D=4$ Dirac matrices
the complex Weyl
to real Majorana representation.

The action \p{actiontw}
produces only the second class constraints
\begin{equation}
 \label{StwB12}
P^{(\l )}_A - \mu_{A}= 0, \qquad  P^{(\mu )}_A = 0,
\end{equation}
\begin{equation}
 \label{StwB34}
\bar{P}^{(\l )}_{\dot{A}} - \mu_{\dot{A}}= 0, \qquad
\bar{P}^{(\mu )}_{\dot{A}} = 0,
\end{equation}
\begin{equation}
 \label{StwF}
\pi^{(\zeta ) } = i \zeta
\end{equation}
The Dirac brackets for the $OSp(8|1)$ supertwistor coordinates are
\begin{equation}
 \label{PBtwB}
[ \mu_{A},\l^B  ]_D = \d_A^{~B},
\qquad [ \bar{\mu}_{\dot{A}},\bar{\l}^{\dot{B}}  ]_D =
\d_{\dot{A}}^{~\dot{B}},
\end{equation}
\begin{equation}
 \label{PBtwF}
\{ \zeta , \zeta \}_D = -i
\end{equation}
They can be also obtained  after the analysis of
the Hamiltonian system described by the original action \p{action}.
For this result one should firstly perform  gauge fixing for all the gauge symmetries,
 arriving at the dynamical  system which contains only second class 
constraints,
 and then pass to the Dirac brackets
in a proper way (see \cite{SorG} for corresponding analysis
of the BS superparticle model).
This means that the generalization of the Penrose correspondence
\p{muXZ2}, \p{bmuXZ2}, \p{zetaTh2} should be regarded as coming from the
 second class constraints (primary and obtained from the gauge fixing)
 of the original system and, thus,
should be considered as a relations hold in the strong sense
(i.e. as operator identities after quantization) \cite{Dirac}.
Hence, after the quantization performed in the frame of supertwistor approach,
the generalized Penrose relations \p{muXZ2}, \p{bmuXZ2}, \p{zetaTh2}
can be substituted into the wave function
in order to obtain
the $D=4$ superspace description of our quantum system.

\bigskip

We shall discuss now the relation of Eq. \p{muXZ2}, \p{bmuXZ2}, \p{zetaTh2},
\p{actiontw} with the known FS $SU(2,2|1)$ supertwistor description
of the BS superparticle \cite{F78,S83,BBCL,stv,SorG,Town91}.
The standard FS description is given by the action
 \begin{equation}\label{actionst}
 S=  - \int  \left(
 {\mu}{}^{A\prime}
d\lambda_{A}
  + \bar{\mu}{}^{\dot{A}\prime} \,
 d{\bar{\lambda}}_{\dot{A}}
  + i d{\xi} ~\bar{\xi} \right)
\end{equation}
supplemented by the first class constraint
\begin{equation}\label{U(1)}
 {\mu}{}^{A\prime} \lambda_{A} -
  \bar{\mu}{}^{\dot{A}\prime} \,
 {\bar{\lambda}}_{\dot{A}}
+ 2i \xi
\bar{\xi} = 0
\end{equation}

The $SU(2,2|1)$ supertwistor
$(\l^A, \bar{\mu}'_{\dot{A}}, \bar{\xi } )$,
contains complex Grassmann variable $\xi$
and the supersymmetric  Penrose--Ferber correspondence
is given by
 \begin{equation}\label{muX}
 \bar{\mu}^{\dot{A}\prime } =
 \left(  x^{B\dot{A}}
 - i \Theta^{B} \bar{\Theta}^{\dot{A}}
 \right)
  \lambda_{B}
\end{equation}
 \begin{equation}\label{xiTh}
\xi = \Theta^A
{\lambda}_{A}, \qquad \bar{\xi}=
\bar{\Theta}^{\dot{A}}
  \bar{\lambda}_{\dot{A}}.
 \end{equation}
  
 Comparing Eqs. \p{actionst} -- \p{xiTh}  with our $OSp(8|1)$ supertwistor description
 \p{muXZ2} -- \p{actiontw}
 of the superparticle \p{action} with additional central charge coordinates,
 we note that
 \begin{itemize}
 \item
Besides  additional terms proportional to
tensorial central charge coordinates $z^{AB}$, $\bar{z}^{\dot{A}\dot{B}}$,
there is present in \p{bmuXZ2} the second term quadratic in Grassmann
variables.
This second term, however, does not contribute to
the invariant $\mu^A \l_A$.
\item
In our model  we get
\begin{equation}\label{notU(1)}
 {\mu}{}^{A} \lambda_{A} -
  \bar{\mu}{}^{\dot{A}} \,
 {\bar{\lambda}}_{\dot{A}}
 = 2\l_A \l_B z^{AB} - 2 \bar{\l}_{\dot{A}}
\bar{\l}_{\dot{B}} \bar{z}^{\dot{A}\dot{B}}
+ 2i \Theta^A \l_A \bar{\Theta}^{\dot{A}}\bar{\l}_{\dot{A}}
\end{equation}
i.e. we do not have additional first class  constraint generating $U(1)$ symmetry
(compare  to \p{U(1)} of the standard supertwistor formulation). Thus our action
\p{actiontw} is not singular in distinction to \p{actionst}, where the
first class constraint \p{U(1)} should be taken into account, e.g. by
introducing it into the
action with Lagrange multiplier \cite{Town91}.

\item
The complex Grassmann variable $\xi$ \p{xiTh} of FS formalism
is replaced in our case
by the real one $\zeta$ \p{zetaTh2}. This difference implies that in our
supertwistor formalism the limit
$z^{AB}\rightarrow 0, ~~
\bar{z}^{\dot{A}\dot{B}}\rightarrow 0$
does not reproduce the standard $SU(2,2|1)$ supertwistor formalism.
Indeed, this is not surprising if we  take into account that,
from algebraic point of view,
$SU(2,2|1)$ is not a subsupergroup of $OSp(8|1)$.

\end{itemize}


\bigskip

\section{$D=4$ Rudychev-Sezgin model in spinor representation}
\setcounter{equation}{0}

Recently the most general superparticle model associated with
space--time superalgebra \p{QQZ}
 was proposed  by Rudychev and Sezgin \cite{RS}.
Introducing generalized real superspace
$(X^{\alpha\beta}, \Theta^{\alpha})$
they consider  the following action
\begin{equation}\label{actionRS}
S=\int d\tau L = \int d\tau \left(
P_{\alpha\beta}\,  \Pi_\tau^{\alpha\beta} +
{1\over 2}
e_{\alpha\beta}\,  P^{\alpha\gamma} \, C_{\gamma\delta}
\, P^{\delta\beta}
\right)\, ,
\end{equation}
where  $ \Pi_\tau^{\alpha\beta}= \dot{X}^{\alpha\beta}
-  \dot{\theta}^{(\alpha} \theta^{\beta)}$
\ $(\dot{a}\equiv {d a \over
d\tau})$, $C$ is the charge conjugation matrix and
$e_{\alpha\beta}$ is the set of Lagrange multipliers,
generalizing einbein in the action for standard Brink-Schwarz massless
superparticle \cite{BS}.

Generalized mass shell condition, obtained by varying
$e_{\alpha\beta}$ in \p{actionRS}, takes the form
\begin{equation}\label{BPS}
P^{\alpha\gamma} C_{\gamma\delta} P^{\delta\beta} = 0 \, .
\end{equation}

In \cite{RS} the model \p{actionRS} was applied for
exotic space--times with more then one time--like dimensions
(for $D=4$ there was considered the model with signature $(2,2)$).
However, it can be considered  as well in the frame of one--time physics.
If we choose for
$\a, \b = 1,\ldots 4$ the charge conjugation matrix $C_{\a\b}$
and the Lagrange multiplier $e_{\a\b}$ to be antisymmetric,
we obtain from \p{actionRS}
the $D=(1+3)$-dimensional model.

The Lagrange multiplier $e_{\a\b}= -e_{\b\a}$
and the generalized  momenta $P_{\a\b}=P_{\b\a}$ can be decomposed as
follows
\begin{equation}\label{e-dec}
e_{\alpha\beta} = C_{\alpha\beta} {1 \over 2} (e+ \bar{e}) +
\left( C\, \gamma_{5} \right)_{\alpha\beta}  {1 \over 2} (e- \bar{e})
+ \left( C \, \gamma_{5} \gamma_{\mu} \right)_{\alpha\beta}
e^{\mu}\, ,
\end{equation}
\begin{equation}\label{P-dec}
P_{\alpha\beta} = \left( C\gamma^{\mu} \right)_{\alpha\beta}
P_{\mu} + \left( C \sigma^{\mu\nu} \right)_{\alpha\beta}
Z_{\mu\nu}\, ,
\end{equation}
Using the decomposition \p{e-dec}, \p{P-dec} and
the Weyl spinor notations we can write the action for the general $D=4$ 
Rudychev-Sezgin model as follows
\begin{equation}\label{acRS4D}
\begin{array}{l}
\displaystyle S = \int d\tau L = \int d \tau
\left( \Pi_\tau^{A\dot{B}} P_{A\dot{B}} + \Pi_\tau^{AB}
Z_{AB} + \Pi_\tau^{\dot{A}\dot{B}} Z_{\dot{A}\dot{B}}\right.
 \\
\nonumber \\
\displaystyle
 + e \epsilon^{AB} \left( P_{A\dot{C}} P_{B}^{\ \dot{C}}
- Z_{AC} Z_{B}^{\ C} \right)
+
\bar{e}\epsilon^{\dot{A}\dot{B}}
\left(
P_{C\dot{A}} P^{C}_{\ \dot{B}} -
Z_{\dot{A}\dot{C}} Z_{\dot{B}}^{\ \dot{C}}\right)
\\
\nonumber \\
\displaystyle
+ ie_{\dot{B}}^{\ A}
\left( Z_{AC}P^{C\dot{B}} - P_{A\dot{C}} Z^{\dot{B}\dot{C}}
\right)\, ,
 \end{array}
\end{equation}

The fermionic constraints are identical with the ones present in our model
\p{DA=0}, \p{bDA=0} and the bosonic constraints are
given by Eq. \p{BPS}, which in the Weyl spinor notation reads
\begin{equation}\label{BPSW}
\begin{array}{l}
\displaystyle
P_{A\dot{B}} P^{A\dot{B}} = Z_{AB} Z^{AB}
= Z_{\dot{A}\dot{B}} Z^{\dot{A}\dot{B}} \, ,
\\
\nonumber \\
\displaystyle
Z_{AC} P^{C\dot{B}} = P_{A\dot{C}} Z^{\dot{B}\dot{C}}\, ,
\end{array}
\end{equation}

The spinorial formulation of the Rudychev--Sezgin model can be obtained
by expressing $P_{A\dot{B}},
Z_{AB}$ and $\bar{Z}_{\dot{A}\dot{B}}$ in terms of spinor coordinates.

Using the technique of spinor Lorentz harmonics \cite{IB90},
one can  show that the general solution of the
constraints \p{BPSW} has the form
\begin{equation}\label{BPSsol}
\begin{array}{l}
\displaystyle
P_{A\dot{B}} = \lambda_{A}\bar{\lambda}_{\dot{B}}
+ u_{A} \bar{u}_{\dot{B}}\, ,
\\
\nonumber \\
\displaystyle
Z_{AB} = Z\lambda_{A} \lambda_{B}
+ \bar{Z} u_{A} u_{B}
\mp i \left( \lambda_{A}u_{B} + \lambda_{B}u_{A}\right)
\sqrt{|Z|^{2} - 1}\, ,
\\
\nonumber \\
\displaystyle
Z_{\dot{A}\dot{B}} = \bar{Z}\bar{\lambda}_{\dot{A}}
\bar{\lambda}_{B}
+ {Z} \bar{u}_{\dot{A}} \bar{u}_{\dot{B}}
\pm i \left( \lambda_{\dot{A}}u_{\dot{B}}
+ \lambda_{\dot{B}}u_{\dot{A}}\right)
\sqrt{|Z|^{2} - 1}\, .
\end{array}
\end{equation}
It is easy to see that
\begin{equation}
P_{A\dot{B}} P^{A\dot{B}} = M^{2}
= |\lambda_{A} u^{A}|^{2} \, .
\end{equation}
Thus, if $\lambda_{A} u^{A} = 0$ we obtain the
model for massless superparticle.
In such a case, because two Weyl spinors are proportional
$u^A \propto \l^A$, it is sufficient
to consider only one spinor $\l^A$.
In such a case Eq. \p{BPSsol} acquires the form
\begin{equation}\label{BPSsol0}
\begin{array}{l}
\displaystyle
P_{A\dot{B}} = \lambda_{A} \lambda_{\dot{B}}\, ,
\\
\nonumber \\
\displaystyle
Z_{AB} = Z \ \lambda_{A} \lambda_{B} \, ,
\qquad
Z_{\dot{A}\dot{B}}
= \bar{Z} \lambda_{\dot{A}} \lambda_{\dot{B}}\, ,
\end{array}
\end{equation}
what leads to the conditions
\begin{equation}
P_{\mu} P^{\mu} = M^{2} = 0\, ,
\qquad
Z_{AB} Z^{AB} = Z_{\dot{A}\dot{B}}
Z^{\dot{A}\dot{B}} = 0 \, ,
\end{equation}
and  the covariant constraints \p{BPSW} are certainly satisfied.

Our model \p{action} appears when $Z=1$, while the standard FS
model corresponds to $Z=0$.
It can be shown that if $Z\not= 1$, there are two fermionic first class constraints,
and thus the model possesses two fermionic gauge symmetries
($\kappa$--symmetries). Only if $Z=1$ we
arrive at the model with three $\kappa$-symmetries, which, in the
brane language, corresponds to the preservation of
$N=3/4$ target space supersymmetries.

\bigskip

\section{Higher--dimensional Generalization.}
\setcounter{equation}{0}

It is quite interesting to consider
a generalization of
our model to $D>4$.
For any $D$ the extension of our generalization of the Cartan-Penrose representation
\p{CPrep} with one $D$--dimensional bosonic spinor $\l_A$ looks as follows
\begin{equation}\label{CPrepD}
 P_{\a\b} = \l_{\a} \l_{\b}, \qquad (\l_\a)^* = \l_\a, \qquad
 \a = 1,\ldots 2^k. \qquad
 \end{equation}
 where \p{CPrep} is obtained  if $k=2$.
 The expression \p{CPrepD} solves the BPS condition
 $det P_{\a\b}=0$ as well as  more strong Rudychev-Sezgin
 BPS constraint \p{BPS} valid in the model \p{actionRS} with antisymmetric charge
conjugation matrix $C~$ ($ C_{\a\b} = - C_{\b\a}$).

Using \p{CPrepD} we get the
multidimensional generalization of our action \p{action}
which reads

\begin{equation}\label{actionD}
S = \int_{{\cal M}^1} \l_{\a} \l_{\b} \Pi^{\a\b}
\end{equation}
$$
 \Pi^{\a\b} = dX^{\a\b} - i d\Theta^{(\a}\Theta^{\b )}, \qquad 
$$
$$
\a =1,...,2^k
$$
and for $k=2$ we get the action \p{action}.

The case
$k=4$ can be treated as describing spinorial
$D=10$ massless superparticle model with $126$ composite tensorial central
charges
$Z_{m_1 \ldots m_5}$ (cf. with \cite{HP,ES}). Indeed, using
the basis of antisymmetric products of $D=10$  sigma matrices
we obtain
\begin{equation}\label{PC10dec}
\l_{\a} \l_{\b} \equiv P_{\a\b} =
P_m \s^m_{\a\b}+
Z_{m_1...m_5}
\s^{m_1...m_5}_{\b\a} ,
  \end{equation}
Contraction of this equation with
$\tilde{\s}^{m\a\b}$ produces the expression for momenta in terms
of bosonic spinors
\begin{equation}\label{Pm10}
P_m = {1 \over 16} \l_{\a} \s_m^{\a\b}\l_{\b}    \qquad
\Rightarrow \qquad P_mP^m = 0.
\end{equation}
The mass shell condition $P_mP^m=0$ appears then as
a result of the $D=10$ identity $(\s_m)_{(\a\b} (\s^m)_{\g ) \d}=0$.

\bigskip

The action \p{actionD} for $k=8$ can be treated as describing a
$0$--superbrane model
in $D=11$ superspace with
$517$ composite tensorial central charge
described by $32$ components of one real Majorana $D=11$ bosonic
spinor.
In distinction to the above case such model
does not produce  a massless superparticle
\footnote{Note, that the $D=11$ Green--Schwarz superparticle model does exist
and was presented in \cite{BT}}.
Indeed, decomposing \p{CPrepD}
in the basis of
products of $D=11$ gamma matrices, one gets
\begin{equation}\label{PC11D}
\l_{\a} \l_{\b} = P_m \G^m_{\a\b}+
Z_{{m}_1{m}_2} \G^{{m}_1{m}_2}
_{\b\a}+ Z_{m_1...m_5}
\G^{m_1...m_5}_{\b\a} ,
\end{equation}
The $D=11$ energy-momentum vector is then given by
\begin{equation}\label{P11lGl}
P_m = {1 \over 32} \l_{\a} \G^{m\a\b}\l_{\b}
\end{equation}
and the $D=11$ mass-shell condition reads
\begin{equation}\label{0P11P11}
M^2 = P_m P^m = {1 \over 1024} (\l \G^{m}\l ) ~(\l \G^{m}\l )
\end{equation}
Using the $D=11$ Fierz identities
one can prove that the mass shell condition acquires the form
\begin{equation}\label{P11P11}
M^2= P_m P^m = 2 ~Z^{mn} Z_{mn} - 32 ^. 5! ~ Z^{m_1...m_5}
Z_{m_1...m_5}
\end{equation}
with $
 Z_{mn} = - {1 \over 64}\l \G_{mn}\l
$,
$~~
 Z_{m_1...m_5} =  {1 \over 32^.5!}\l \G_{m_1...m_5}\l
$.

If we take into consideration that
the equations of motion for our model \p{actionD} imply that
the bosonic spinor $\l^\a$ is constant  ($d\l^\a=0$), we have to conclude that
\p{actionD} with $k=8$ provides the $D=11$ superparticle model
with mass generated dynamically in a way similar to the
tension generating mechanism, studied in
superstring and higher branes in \cite{generation}.

                   \bigskip

Performing the integration by parts we can rewrite the action \p{actionD}
in the $OSp(1|2^k)$ (i.e. $OSp(1|16)$ for $D=10$ and $OSp(1|32)$
for $D=11$) supertwistor $Y^{{\cal A}}= (\mu^\a, \zeta )$ components:
\begin{equation}\label{actiontwD}
S= - \int (\mu^\a d\l_\a + i d\zeta ~\zeta ), \qquad \a = 1,\ldots , 2^k.
\end{equation}
The generalized Penrose--Ferber  correspondence
between real supertwistors and real generalized superspace looks as follows
\begin{equation}\label{muXZD}
\mu^\a = X^{\a\b} \l_\b - i \Theta^\a (\Theta^\b \l_\b), \qquad
\zeta = \Theta^\a \l_\a .
\end{equation}

More detailed discussion of higher dimensional case will be given in our subsequent publication.

\section{Final remarks}

In this paper we proposed
and discussed in some detail
a new $D=4$ massless
superparticle model with three $\kappa$--symmetries.
These $\kappa$-symmetries
 correspond to target space supersymmetries preserved by the
 BPS configuration. The BPS configurations are identified usually with some
 supersymmetric branes
 or their intersections  \cite{Bars,PKT,Hewson,SorT}.
 The triviality of realization of a
 part of target space supersymmetry is explained
 by the presence of the
 corresponding number of  fermionic gauge $\kappa$--symmetries.
Thus our case  corresponds to BPS configurations preserving
$3/4$ of the target space supersymmetry.
Such configurations were not known before, as the usual superbranes
conserve not more then $1/2$ supersymmetries, while their intersections
keep $1/4$ and less supersymmetries.

                  \bigskip

We would like also recall that
in the  'M-theoretic' approach (see e.g. \cite{AT,PKT,SorT})
the tensorial central charges $Z_{m_1\ldots m_p}$ are
considered as carried by p-branes.
 Following such treatment, one should interpret e.g.  in $D=4$ central charges
 $Z_{\mu\nu}$ as an indication of presence of $D=4$ supermembrane ($p=2$).
 The relation of our superparticle model with such $D=4$ membrane states is not clear now and
 can be regarded as an interesting subject for further study.
Here we should only guess that there should be some singular point--like limit
of supermembrane, which should keep the nontrivial topological charge
and increase the number of preserved (realized linearly)
$D=4$ target space supersymmetries.
Similar limiting prescription should be possible
e.g. for 5--branes in $D=10, 11$ leading to the $D=10$ and $D=11$
superparticle actions
\p{actionD} with the relation \p{CPrepD}
describing composite tensor charges.

At the end of the paper (see \p{actiontwD}, \p{muXZD}) we only proposed a
generalized FS model for $D>4$. We shall consider in more detail the cases
of $D=10$ and $D=11$ (as well as $D=12$ with two times \cite{Bars,RS})
in the nearest future.

\bigskip

\centerline{{\bf Acknowledgements}}

\bigskip

The authors would like to thank D. Sorokin for useful discussions and 
E. Sezgin and I. Rudychev for interest to the work.
One of the author (I.B.) is grateful for the hospitality at the
Institute for Theoretical Physics of Wroclaw University where the present paper
was completed.

The work was supported by the Austrian Science Foundation in the form
of the Lise Meitner Fellowship under the Project
 {\bf M472--TPH}, by the INTAS grant {\bf INTAS-96-308} and 
{\bf KBN} grant {\bf 2P03B13012}.

\end{document}